\documentclass[11pt,superscriptaddress,notitlepage,longbibliography]{revtex4-1}
\usepackage{amsmath,amssymb,amsthm}
\usepackage{graphicx}
\usepackage{hyperref}
\newcommand{\p}{\partial}
\begin{document}
\title{Fluid flow structures in an evaporating sessile droplet depending on the
droplet size and properties of liquid and substrate}
\author{M.N. Turchaninova}
\affiliation{National Research University Higher School of Economics, 101000 Moscow, Russia}
\affiliation{Landau Institute for Theoretical Physics, 142432 Chernogolovka, Russia}
\author{E.S. Melnikova}
\affiliation{National Research University Higher School of Economics, 101000 Moscow, Russia}
\affiliation{Landau Institute for Theoretical Physics, 142432 Chernogolovka, Russia}
\author{A.A. Gavrilina}
\affiliation{National Research University Higher School of Economics, 101000 Moscow, Russia}
\affiliation{Landau Institute for Theoretical Physics, 142432 Chernogolovka, Russia}
\author{L.Yu. Barash}
\affiliation{Landau Institute for Theoretical Physics, 142432 Chernogolovka, Russia}
\begin{abstract}
We investigate numerically quasi-steady internal flows
in an axially symmetrical evaporating sessile droplet depending on 
the ratio of substrate to fluid thermal conductivities, fluid volatility,
contact angle and droplet size.
Temperature distributions and vortex structures are obtained
for droplets of 1-hexanol, 1-butanol and ethanol.
To this purpose, the hydrodynamics of an evaporating sessile drop, 
effects of the thermal conduction in the droplet and substrate 
and diffusion of vapor in air have been jointly taken into account.
The equations have been solved by finite element method using ANSYS Fluent.
The phase diagrams demonstrating the number and orientation 
of the vortices as functions of the contact angle and the ratio of substrate 
to fluid thermal conductivities, are obtained and analyzed for various values of parameters.
In particular, influence of gravity on the droplet shape
and the effect of droplet size have been considered.
We have found that the phase diagrams of highly volatile droplets
do not contain a subregion corresponding to a reversed single vortex,
and their single-vortex subregion becomes more complex.
The phase diagrams for droplets of larger size 
do not contian subregions corresponding to a regular single vortex 
and to three vortices. 
We demonstrate how the single-vortex subregion disappears with a gradual increase
of the droplet size.
\end{abstract}
\maketitle

\section{Introduction}

Sessile drop evaporation processes and structures of the fluid flow are of interest 
for important applications in science, engineering and medicine.
The structure of fluid flow produced by thermocapillary phenomena inside an evaporating droplet 
is important in many of the applications and has been
intensively studied~\citep{brutin2018,erbil2012,larson2014,thiele2014,tarasevich2019}.
In particular, the Marangoni convection in a droplet
is usually of importance in such applications as
evaporative lithography~\citep{harris2007,KolegovBarash2020} 
and self-assembly of nanocrystal superlatices~\citep{bigioni2006}.
Axisymmetric Marangoni flows can often 
be observed at room temperature and under atmospheric 
pressure~\citep{Savino2004,Kang2004,HuLarsonReversal,ristenpart2007}.

The substrate properties play an important role in forming the hydrodynamic flow structure.
It was originally observed and described in \citep{ristenpart2007} that the circulation direction 
close to the contact line depends on the substrate thermal conductivity.
An alternative approach by Xu et al.~\citep{xu2010} focused on
a heat transfer in the immediate vicinity of the symmetry axis piercing the apex. 
The transition between the opposite 
circulation directions, taking place with a variation of the relative 
substrate--liquid thermal conductivity, has been identified in~\citep{xu2010}
under the conditions that differ from those obtained in Ref.~\citep{ristenpart2007}.
Hu and Larson demonstrated that fluid circulation
in the vortex can reverse its sign at a critical contact angle 
for a drop placed onto substrates with finite thicknesses~\citep{HuLarsonMarangoni}.
The influence of substrate temperature has been studied
in a number of experiments~\citep{patil2016,girard2010,sobac2012}.

The thermal conduction processes throughout the droplet
can generally result in a nonmonotonic spatial dependence of the surface 
temperature and in more complicated convection patterns inside a drop.
The temperature distribution can be qualitatively
understood as a result of matching the
heat transfer through the solid-liquid and liquid-vapor interfaces.
With varying relative substrate--liquid thermal conductivity and/or contact angle,
transitions between regimes with different numbers of vortices and/or
circulation directions take place~\citep{zhang2014,barash2015}.

In order to form the vortex structure inside the droplet in a controlled manner, 
one should know the dependence of the flow patterns on such
externally varying parameters as thermal conductivities of liquid and substrate,
substrate thickness, liquid volatility, contact angle and droplet size.
Here we study such a dependence by performing the numerical simulation 
of Marangoni convection in a drop.
The obtained results extend those of Ref.~\citep{barash2015},
which mainly focus on the substrate thickness but did not
consider the dependence of fluid flows on the liquid volatility and droplet size.

\section{Basic equations}
\label{eqSec}

\begin{figure}[t]
\begin{center}
\includegraphics[width=0.6\textwidth]{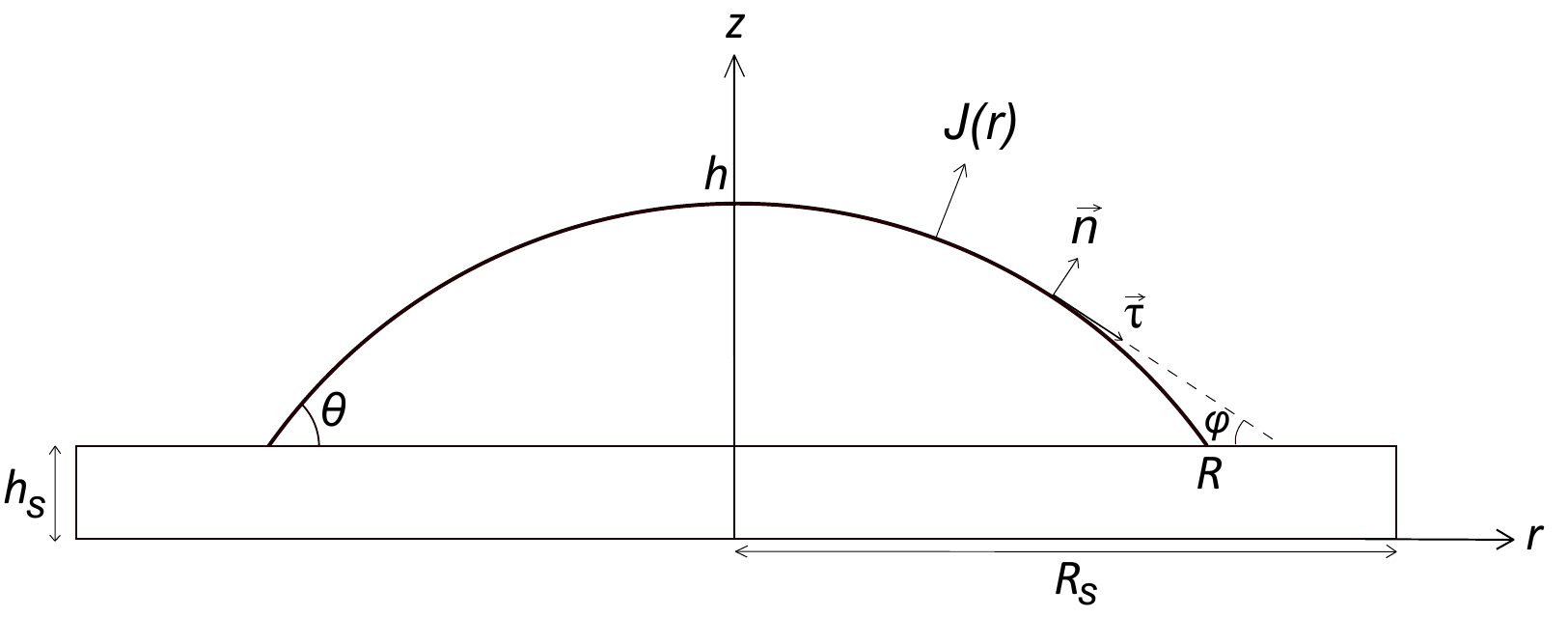}
\end{center}
\caption{
Sessile droplet on a solid substrate. The symmetry axis is $z$, the contact angle is $\theta$, and the local evaporation flux is $J(r)$.
}
\label{dropletfig}
\end{figure}

We consider axially symmetrical evaporating sessile droplet 
with a contact line pinned to a hydrophilic solid substrate (Figure \ref{dropletfig}).
The droplet shape can be described with good accuracy within a spherical cap approximation
when the Bond number $Bo = \rho g h_0 R/(2 \sigma \sin{\theta})$ and the capillary number
$Ca = \eta \overline{v}/\sigma$ are much smaller than unity 
(see the notations in Table~\ref{table1}).

The axial symmetry of thermocapillary fluid flow can break down for a strong evaporation 
of a substantially heated droplet
and a resulting strong fluid velocity~\citep{Sefiane2008,Carle2012,Shi2019}.
However, axisymmetric flows can usually be observed at room temperature and under atmospheric
pressure~\citep{ristenpart2007}.

The dynamics of vapor concentration $u(r,z)$ in the surrounding atmosphere is described by the diffusion equation
\begin{equation}\label{diff}
\dfrac{\partial u}{\partial t} = D \Delta u.
\end{equation}
The boundary conditions are: $u = u_s$ on the drop surface, $u = 0$ far away from the drop, 
$\partial u / \partial r = 0$ and $\partial u / \partial z = 0$ on the axes $r = 0$ and $z = 0$, correspondingly.

In the quasi-stationary approximation, the diffusion equation~(\ref{diff}) can be replaced 
by the Laplace's equation $\Delta u = 0$.
The approximation can be employed if the vapor concentration adjusts on time scales 
much less than the total drying time (i.e., $R^2/D$ is much smaller than $t_f \approx 0.2 \rho R h_0/(D u_s)$),
and if the Stefan flow can be neglected. The latter condition holds at room temperature and under atmospheric pressure,
when the saturated vapor concentration is much smaller than the density of air~\citep{Fuchs1959}.
Within the quasi-stationary approximation and the spherical cap approximation, 
the analytical solution of the diffusion equation can be obtained which gives the inhomogeneous evaporation flux from
the surface of the evaporating droplet~\citep{deegan2000}
\begin{equation}
J_s(r) = \dfrac{D u_s}{R} \bigg(\dfrac{\sin{\theta}}{2} + \sqrt{2} (x(r) + \cos{\theta})^{3/2}\displaystyle\int_{0}^{\infty} \dfrac{\cosh{(\theta\tau)}}{\cosh{(\pi \tau)}} \tau \tanh((\pi - \theta)\tau) P_{-1/2+i\tau} (x(r)) d\tau\bigg),\label{analytical}
\end{equation}
where 
\begin{equation}
x(r) = \bigg(r^2 \cos{\theta}/R^2 + \sqrt{1-r^2 \sin^2{\theta} / R^2}\bigg) \bigg\slash (1-r^2/R^2)
\end{equation}
and $P_{-1/2+i\tau} (x)$ is the Legendre polynomial. 
Expr.~(\ref{analytical}) can be conveniently 
approximated with high accuracy as~\citep{deegan2000}

\begin{equation}\label{Jr}
J_s (r) = J_0(\theta) (1-r^2/R^2)^{-(1/2-\theta/\pi)},
\end{equation}
where $J_0 (\theta)$ can be determined by the following expressions~\citep{hu2002}:

\begin{eqnarray}
J_0 (\theta) / (1-\Lambda(\theta)) &=& J_0(\pi/2) (0.27 \theta^2 + 1.3),\\
\Lambda(\theta) &=& 0.2239(\theta-\pi/4)^2+0.3619,\\
J_0(\pi/2) &=& D u_s/R.
\label{J0}
\end{eqnarray}

\begin{table}[t]
\caption{The notations and the parameter values used in the calculations.}
\label{ParamTable}
\begin{center}
\small
\begin{tabular}{cllllll}
\hline
&&&& ethanol & 1-butanol & 1-hexanol \\
\hline
Drop & Initial temperature & $T_0$ & K  & $293.15$  & $293.15$ & $293.15$ \\
parameters & Contact line radius & $R$ & cm   & $0.1$ & $0.1$ & $0.1$ \\
\hline
Fluid &Density & $\rho$ & g/cm$^3$    & $0.789$ & $0.8098$ & $0.8136$ \\
parameters &Molar mass & $\mu$ & g/mole & $46.07$ & $74.122$ & $102.17$\\
&Thermal conductivity & $k$ & W/(cm$\cdot$ K) & $1.69\cdot 10^{-3}$  & $1.54\cdot 10^{-3}$ & $1.50\cdot 10^{-3}$  \\
&Heat capacity  & $c_p$ & J/(mole K)   & $112.3$ & $177.2$ & $240.4$  \\
&Isochoric heat capacity & $c_v$ & J/(g K) & $1.787$ & $1.83$ & $1.889$ \\ 
&Thermal diffusivity & $\kappa$ & cm$^2$/s & $8.79\cdot 10^{-4}$ & $7.95\cdot 10^{-4}$ & $7.84\cdot 10^{-4}$ \\
&Dynamic viscosity & $\eta$ &  g/(cm$\cdot$ s) & $1.074\cdot 10^{-2}$ & $2.544\cdot 10^{-2}$ & $4.578\cdot 10^{-2}$ \\
&Surface tension & $\sigma$ &  g/s$^2$  & $21.97$ & $24.93$ & $25.81$ \\
&$-\p($surface tension$)/\p T$ & $-\sigma'_T$ & g/(s$^2\cdot$ K) &$0.0832$ &$0.0898$ &$0.08$ \\
&Latent heat of evap. & $L$ & J/g & $918.6$ & $706.27$  & $603.0$  \\
\hline
Substrate & Radius & $R_S$ & cm & $0.125$ & $0.125$ & $0.125$ \\
parameters & Thickness & $h_S$ & cm & $0.02$ & $0.02$ & $0.02$ \\
\hline
Vapor &Diffusion constant & $D$ & cm$^2$/s  & $0.1181$ & $0.0861$ & $0.0621$ \\
parameters &Saturated vapor density & $u_s$ & g/cm$^3$ & $1.46\cdot 10^{-4}$ & $2.76\cdot 10^{-5}$ & $6.55\cdot 10^{-6}$ \\
\hline
\end{tabular}
\end{center}
\label{table1}
\end{table}

The above model of diffusion-limited evaporation assuming a constant vapor pressure along the interface
is valid for a relatively small temperature gradients and/or only a minor slope in the vapor pressure chart.
For a particular cases considered in the present study, the temperature difference along the interface is
found to be of the order of $1$~K, and there is $\sim 5\%$ relative deviation of saturated vapor density 
as a result of the $1$~K change in temperature. Therefore, the above model can be employed to a first approximation.

It follows from~(\ref{Jr}) that the evaporating flux density is inhomogeneous along the surface
and increases substantially on approach to the pinned contact line.
A resulting inhomogeneous mass flow and corresponding heat transfer 
modify the temperature distribution along the droplet surface and, therefore,
result in appearence of Marangoni forces associated with the temperature-dependent surface tension,
which, in turn, generate convection inside the droplet.

The basic hydrodynamic equations inside the drop are the Navier-Stokes equations  
and the continuity equation for the incompressible fluid
\begin{equation}
\label{NS}
\frac{\partial \textbf{\textbf{v}}}{\partial t} + (\textbf{\textbf{v}} \cdot \nabla) \textbf{\textbf{v}} + \frac{1}{\rho} \text{ grad } p = \nu \Delta \textbf{\textbf{v},}
\end{equation}

\begin{equation}
\text{div \textbf{v}} = 0.
\end{equation}

Here $\Delta = \partial^2 / \partial r^2 + \partial / r \partial r + \partial^2 / \partial z^2$, 
$\nu = \eta / \rho$ is kinematic viscosity, $\textbf{v}$ is fluid velocity, $p$ is pressure.
The boundary conditions are: $\left(v_r\right)_{r=0} = 0$, $\left(\partial v_z/\partial r\right)_{r=0} = 0$,
$\textbf{v} = 0$ at the solid--liquid interface (at $z=h_S$) and 
$\partial \sigma / \partial \tau = \eta (\partial v_{\tau}/ \partial n - v_{\tau} \partial \varphi / \partial \tau)$ 
at the liquid--vapor interface~\citep{barash2009}.

The buoyancy force has been neglected in~Eq.(\ref{NS}), because the buoyancy-induced convection
is much weaker than Marangoni flow when $\rho g h^2\beta/(7\sigma') \ll 1$,
where $\beta$ is thermal expansion coefficient~\citep{Pearson1958}.
The latter condition holds for all the cases considered in the present study,
including droplets of larger size in Sec.~\ref{PhaseDiagramsSec}, where we always have $h < 0.1$~cm.

The calculation of temperature distribution is carried out using the thermal conduction equation

\begin{equation}
\dfrac{\partial T}{\partial t} + \textbf{ v} \cdot \nabla T = \kappa \Delta T.
\end{equation}

For relatively small and slowly evaporating droplets, when $Pe=\overline{v}R/\kappa \ll 1$,
the convective heat transfer is much smaller than conductive heat transfer, 
and, hence, the velocity field does not influence the thermal conduction:
\begin{equation}\label{without_conv}
\dfrac{\partial T}{\partial t}  = \kappa \Delta T.
\end{equation}

This is the case, for example, for the droplet of 1-hexanol (see the parameters in Table~\ref{table1}).
The thermal conduction inside the substrate is also described by (\ref{without_conv}).

We note that the transient time for heat transfer $t_{heat}=Rh_0/\kappa$,
transient time for momentum transfer $t_{mom}=\rho Rh_0/\eta$ and
transient time for vapor phase mass transfer $t_{mass}=u_s/\rho\cdot t_f$
are much smaller than the total drying time $t_f\approx 0.2\rho R h_0/(D u_s)$.
This permits to describe the quasistationary stage of the evaporation process
assuming a fixed geometric shape and
disregarding the time derivatives in the heat conduction equation, 
Navier--Stokes equation and the diffusion equation~\citep{larson2014}.

Considering a fixed geometric shape, we also disregard the capillary flow,
which is the coffee-stain outward flow~\citep{deegan2000}.
This is justified in presence of a strong recirculating flow
due to the Marangoni effect.
Indeed, the velocity of the capillary flow is of the order of
$R/t_f$ when the contact angle is not too small (see, e.g., the expressions in~\citep{Popov2005,HuLarson2005,KolegovBarash2019}).
Our numerical results confirm that the characteristic velocities of Marangoni flow are orders of magnitude larger than $R/t_f$.

The boundary conditions for the thermal conduction take the form 
$\partial T / \partial r =  0$ for $r = 0$; $T = T_0$ for $z = 0$; 
$\partial T / \partial n = 0$ at the substrate--gas interface,
$k_S \partial T_S / \partial z = k_L \partial T_L / \partial z$ at the substrate--fluid interface,
$\partial T / \partial n = -Q_0(r)/k$ at the drop surface. 
Here $T_L$ and $T_S$ correspond to the liquid and substrate temperature,
$k_L$ and $k_S$ are thermal conductivities of liquid and substrate, respectively,
$Q_0(r) = LJ_s(r)$ is the rate of heat loss per unit area of the upper free surface, 
\textbf{n} is a normal vector to the drop surface, $J_s(r)$ is determined by (\ref{Jr}).

\section{Numerical simulation}
We solve the thermal conduction equation and the Navier-Stokes equations by finite element method 
using computational fluid dynamics simulation package ANSYS Fluent
designed for modeling the laminar flow, turbulence, heat transfer, and reactions 
for industrial applications.

The simulation includes several steps.
\begin{itemize}
\item \textit{Geometry and mesh generation}

We use two-dimensional geometry of the axially symmetrical surface according to Table~\ref{table1}. 
The surface curvature radius and the droplet height are equal to $R/\sin{\theta}$ 
and $R(1/\sin{\theta} - \cot{\theta})$, correspondingly.
In ANSYS Fluent, the symmetry axis should be denoted as $x$-axis
(as distinct from $z$-axis in Fig.~\ref{dropletfig}).
The next step is to divide the simulation region into small computational cells. 

\item \textit{Defining model and setting properties}

We choose the pressure-based solver type, 2D axisymmetric geometry, steady mode for time, viscous (laminar) flow 
and enable the calculation of energy in the model.
We specify all physical properties of the fluid such as viscosity and density.
The values are listed in Table~\ref{table1} including the temperature derivative of the surface tension
$\sigma^{\prime}_{T} = -\partial \sigma / \partial T$ which allows us to specify the properties of Marangoni stress.

\item \textit{Setting boundary conditions and solution method}

The boundary conditions are set according to Section~\ref{eqSec}.
Some of the boundary conditions, such as at the symmetry axis,
are automatically taken into account in ANSYS Fluent.
The rate of heat loss $Q_0 = LJ_s (r)$ is specified with user-defined function (UDF) 
written in the C programming language.
The solution algorithm is SIMPLE (Semi-Implicit Method for Pressure Linked Equations). 
It uses a relationship between velocity and pressure corrections to enforce mass conservation 
and to obtain the pressure field. The algorithm is written in such a way that the continuity equation 
is automatically satisfied. See~\citep{patankar} for detailed description of the SIMPLE algorithm.


\item \textit{Post-processing}

We display the simulation results: velocity field, absolute values of velocity and
temperature distribution, and analyse the vortex structure.

\end{itemize}

We note that our mesh contained at least 70000 cells for each contact angle.
We used at least 20000 iterations to ensure that we obtained a steady state of the fluid flows
in each case.
Benchmark tests have been performed to ensure that 
a further increase of numbers of cells and iterations
does not change the numerical results.
Occasionly, when the obtained fluid flow structure was controversial, 
the numbers of cells and iterations were further increased to ensure the correctness
of the results.

\section{Phase diagrams for the number and orientation of vortices}
\label{PhaseDiagramsSec}

\begin{figure}[t]
\centering
\includegraphics[width=0.48\textwidth]{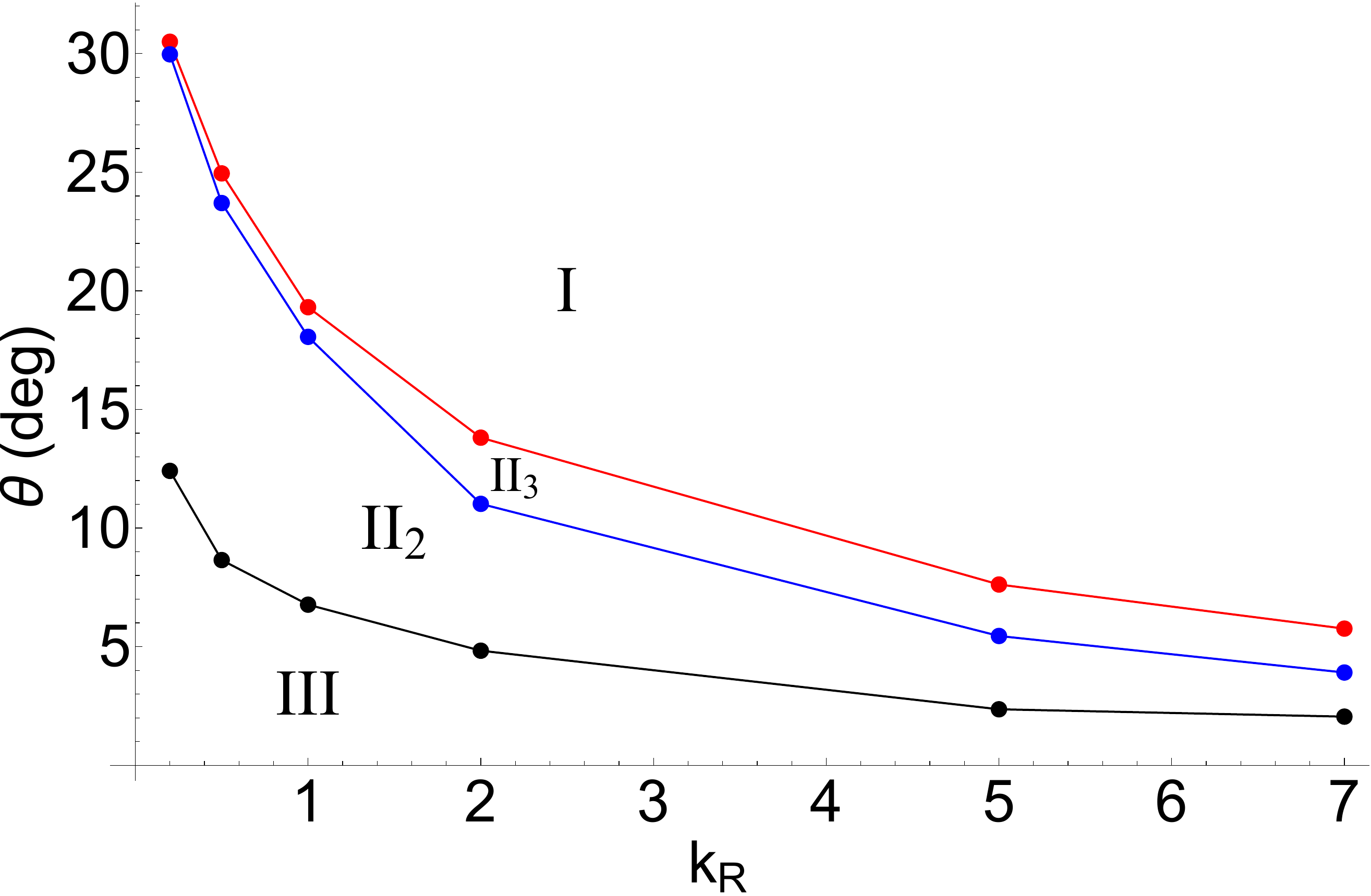}
\includegraphics[width=0.48\textwidth]{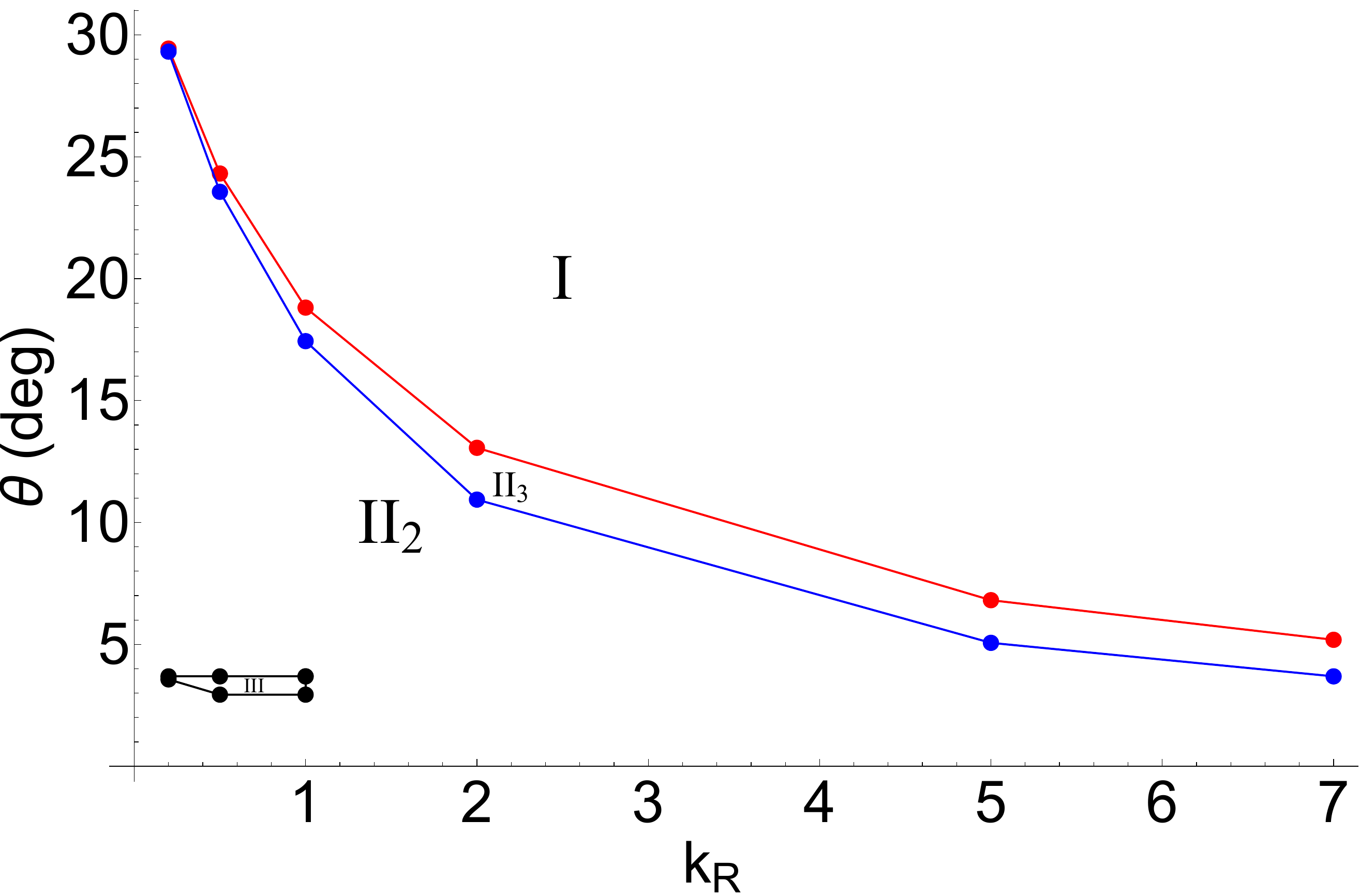}\\
(a)\hspace*{220pt}(b)\\
\includegraphics[width=0.48\textwidth]{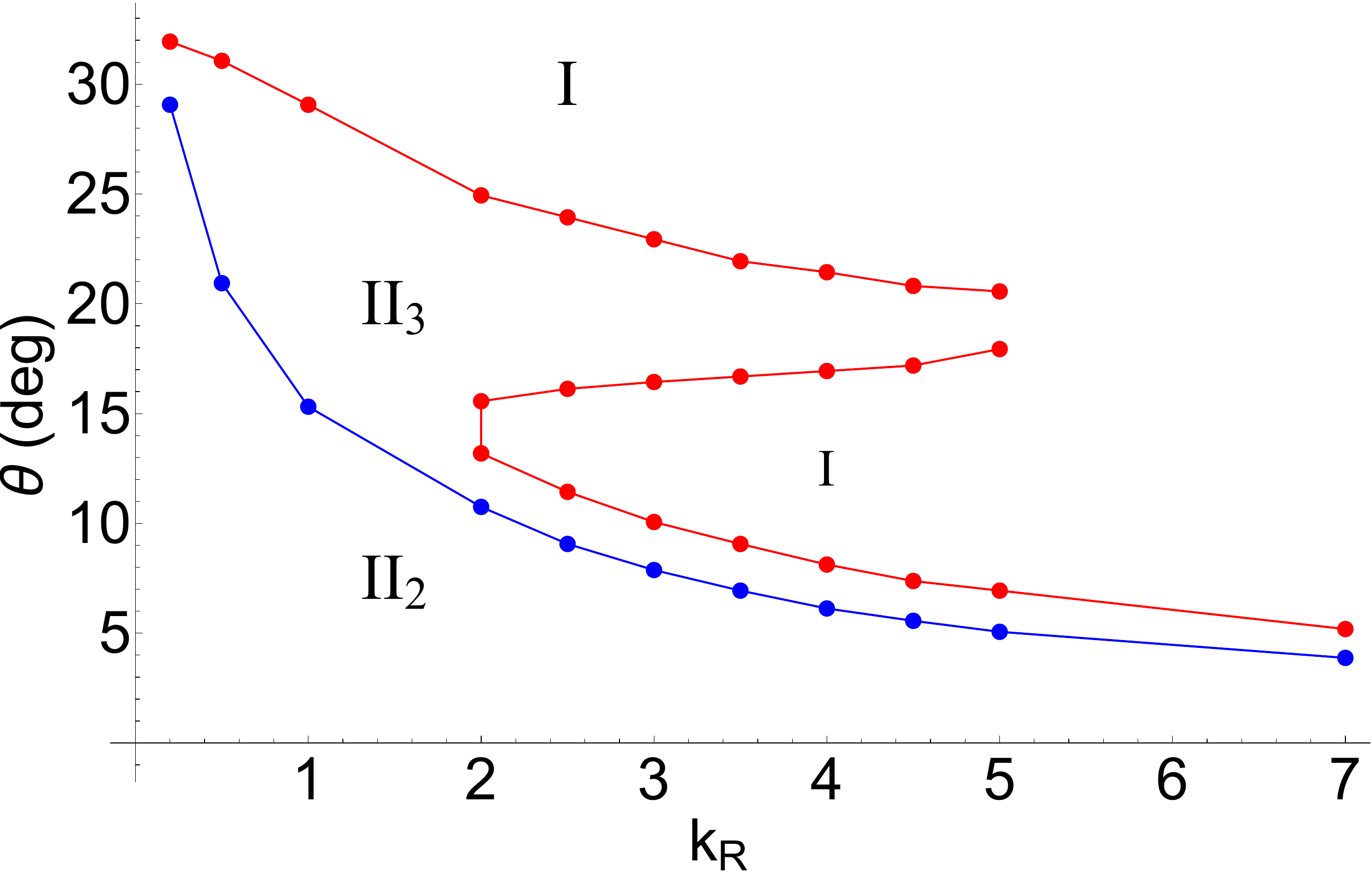}\\
(c)\\
\caption{
Results for $\theta$ vs $k_R = k_S / k_L$ for the droplet of (a) 1-hexanol, (b) 1-butanol and (c) ethanol, where the substrate temperature at the lower boundary is room temperature.}
\label{fig:threeliquids}
\end{figure}

\begin{figure}[t]
\begin{center}
\includegraphics[width=0.7\textwidth]{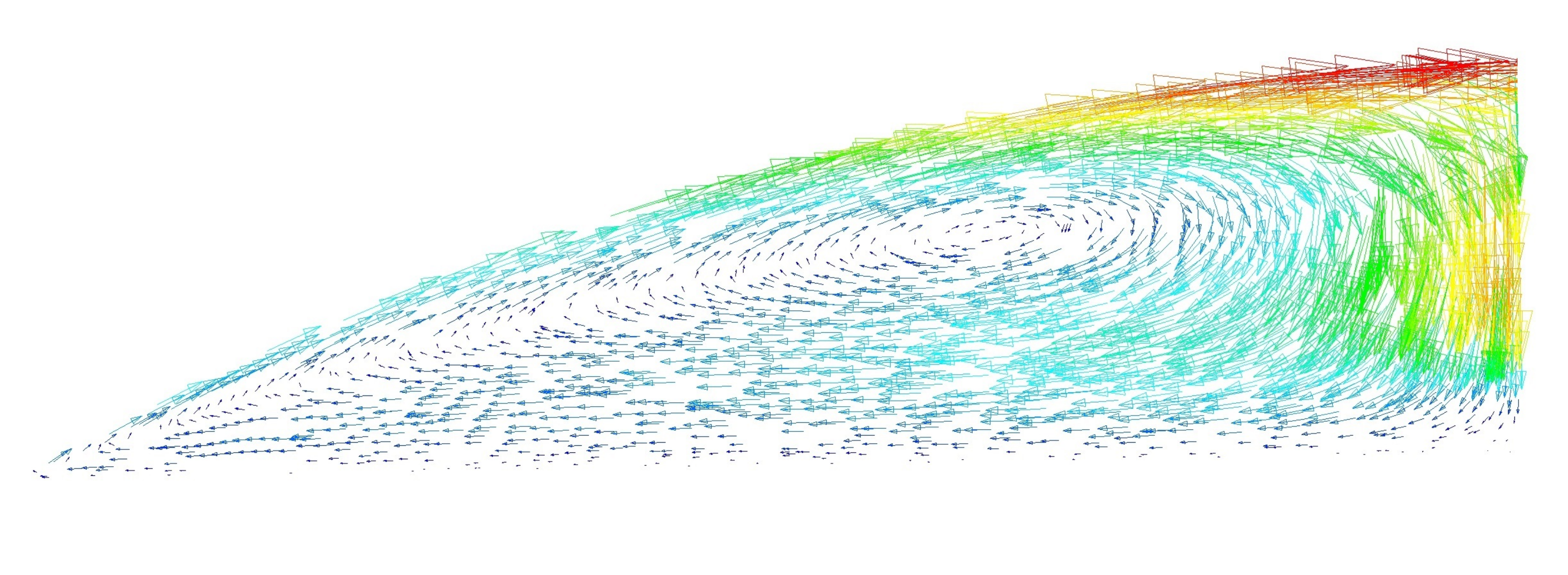}\\
\vspace{-0.6cm}
(a)\\
\includegraphics[width=0.7\textwidth]{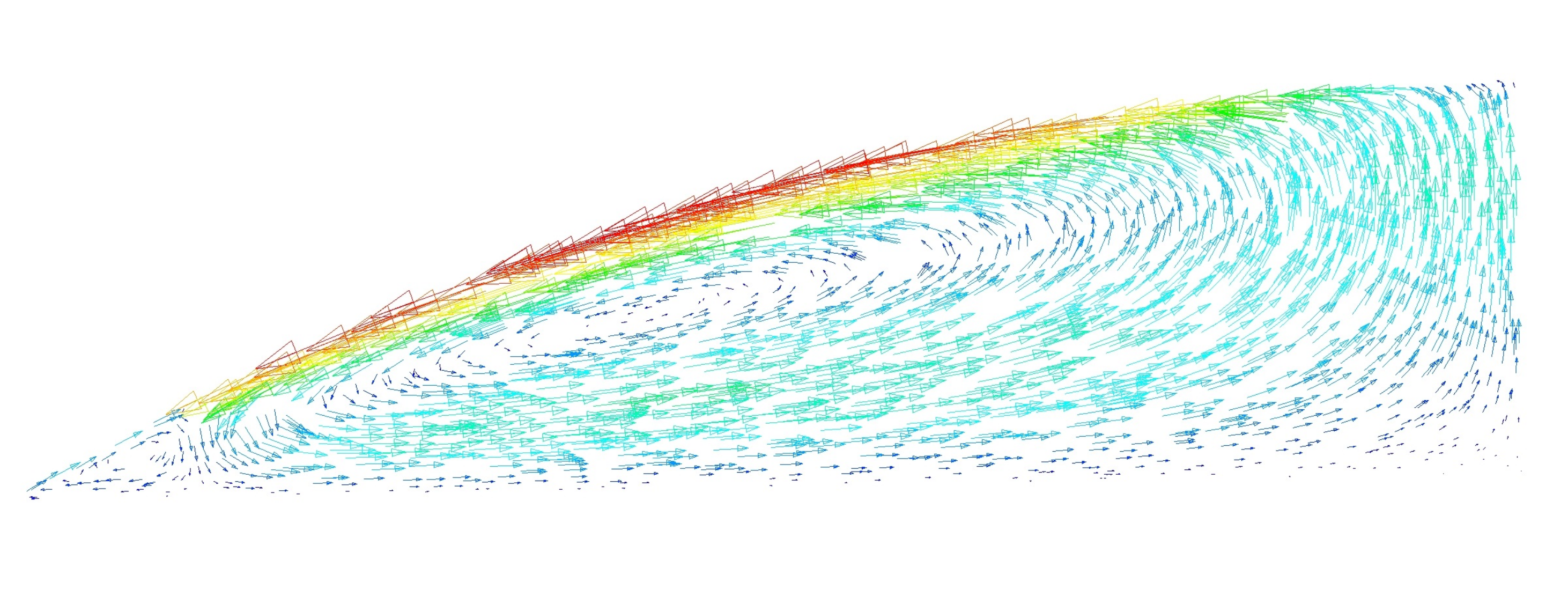}\\
\vspace{-0.6cm}
(b)\\
\includegraphics[width=0.7\textwidth]{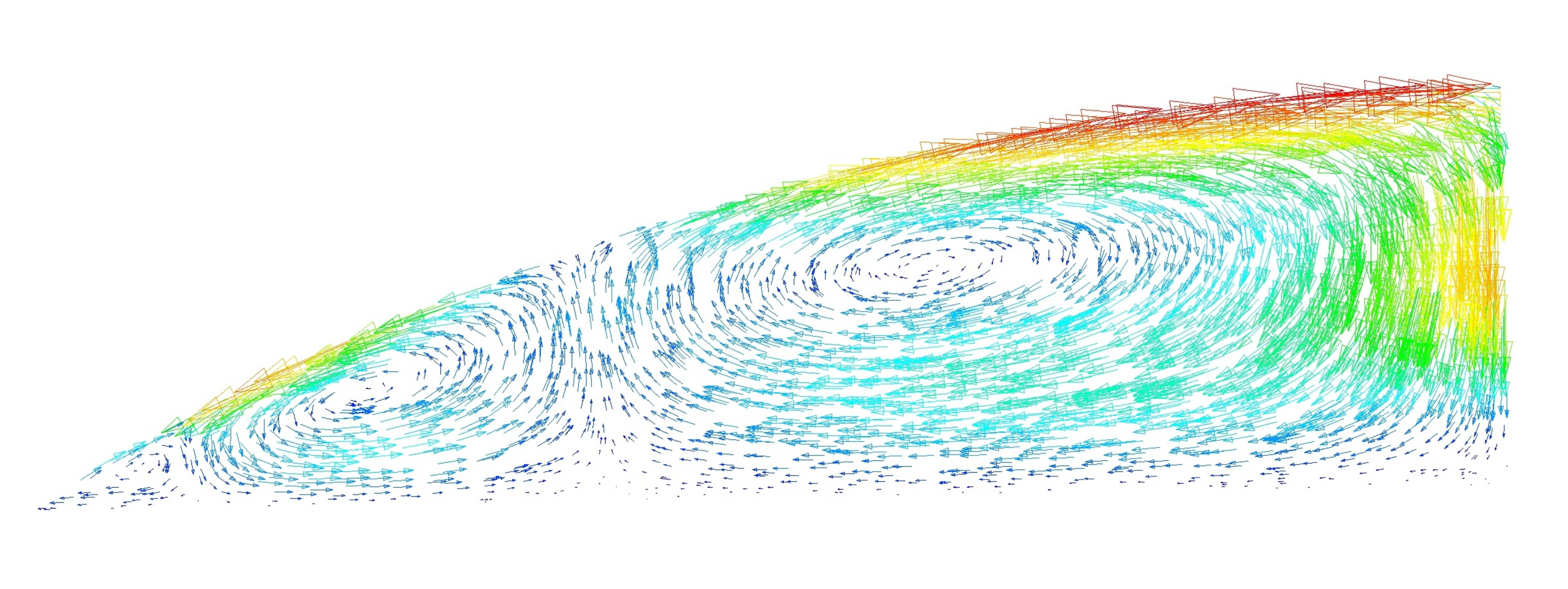}\\
\vspace{-0.6cm}
(c)\\
\includegraphics[width=0.7\textwidth]{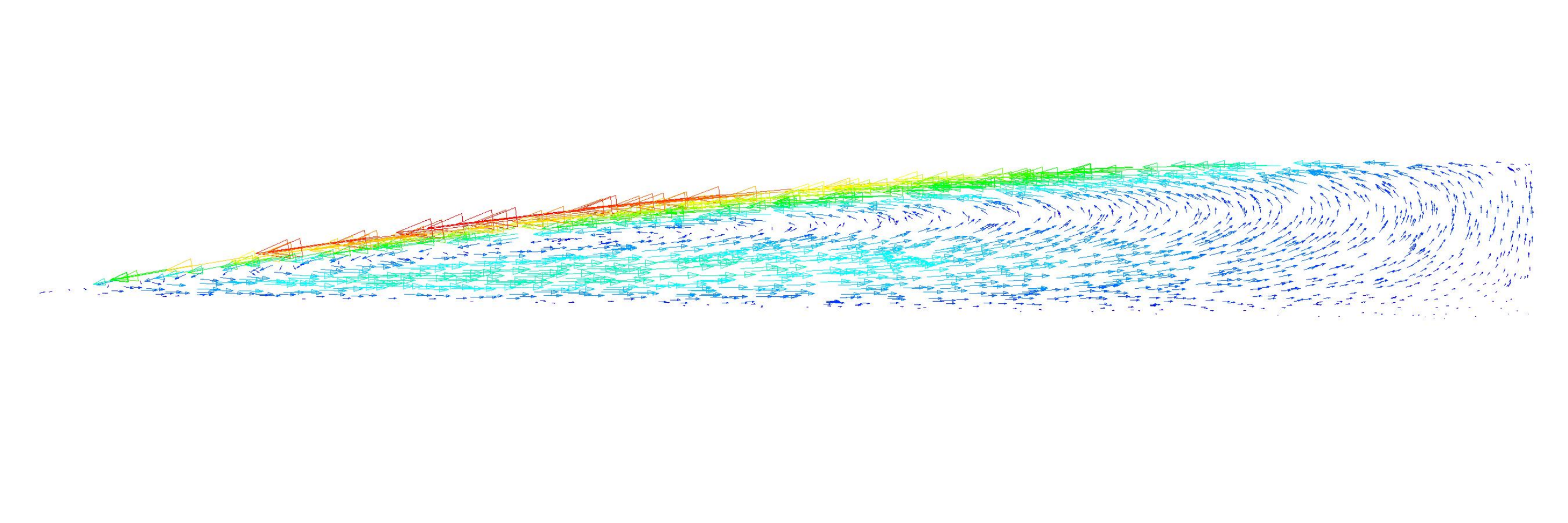}\\
\vspace{-1.5cm}
(d)
\end{center}
\caption{Examples of vector field plots of velocity obtained for (a) single-vortex, 
(b) two-vortex, (c) three-vortex and (d) reversed single vortex regimes.}
\label{fig:vortex}
\end{figure}

Consider the phase diagrams which represent the number and orientation of the vortices 
for quasistationary fluid flows in the droplet as functions of $k_R$ and $\theta$.
Here, $k_R$ is ratio of substrate to fluid thermal conductivities and $\theta$ is the contact angle.
The fluid flow fields have been numerically obtained and different phase
diagram regions identified for several particular examples chosen
(see Fig.~\ref{fig:threeliquids}).
The notations I, II$_2$, II$_3$, III signify the phase diagram regions with
single-vortex, two-vortex, three-vortex and reversed single-vortex regimes,
respectively~(Fig.~\ref{fig:vortex}).

We have determined the phase diagram regions via the numerically obtained fluid flow fields.
The surface temperature distrubition usually follows the vortex structure in the 
droplet, although there are some exceptions to this rule 
due to the inertia of the fluid flow.

In our phase diagrams, the blue curve and black curve correspond to the change of sign 
of the tangential component of temperature gradient 
at the liquid--vapor interface, at the droplet apex and near the contact line, correspondingly.
As shown in Ref.~\citep{barash2015}, with an increase of the substrate thickness,
the subregion II$_2$ becomes dominating over II$_3$.
Further increase of the substrate thickness results in
shifting the regions I and II to larger contact angles.
The blue curve is situated below
the black curve at small values of $h_R=h_S/R$,
while for $h_R \gtrsim 0.05$ the blue curve is above the black curve~\citep{barash2015}.
The phase diagrams in Fig.~\ref{fig:threeliquids} correspond to the latter case.

For $h_R \gtrsim 0.05$,
the transition between the regions III and II$_2$ 
corresponds to the change of sign 
of the tangential component of the temperature gradient at the liquid--vapor interface
near the contact line.
The transition between the regions II$_2$ and II$_3$
corresponds in this case to the change of sign of the tangential component 
of the temperature gradient 
at the liquid--vapor interface at the droplet apex.
The transition between the regions I and II$_3$
corresponds to the change between
a nonmonotonic dependence of surface temperature on $r$
and a monotonically increasing surface temperature with $r$.

The main features of the phase diagram can be qualitatively
understood as a result of matching the heat transfer 
across the solid--liquid interface and the vaporization heat
through the liquid--vapor interface
(see details in~\citep{zhang2014,barash2015}).
As a result of the matching and competition of these two effects, 
vortex structures are formed which depend
on the values of $k_R$ and $\theta$.

Consider now the volatility effect on the phase diagram.
Fig.~\ref{fig:threeliquids} shows the phase diagrams obtained
for the three droplets of different liquids,
where the parameters of the droplets are listed in Table~\ref{table1}.
Since the saturated vapor density of 1-hexanol, 1-butanol and ethanol
differ by orders of magnitude as seen in Table~\ref{table1},
the evaporation rates (volatilities) of these liquids 
described by Eqs.(\ref{Jr})-(\ref{J0}) also significantly differ.

As seen in Fig.~\ref{fig:threeliquids},
with an increase of the fluid volatility,
the subregion III disappears;
the subregion II$_2$ increases at the expense
of the subregion III; 
the subregion II$_3$ becomes larger at the expense
of the subregion I;
the subregion I diminishes for small values of $k_R$
and becomes more complex for highly volatile liquids.

\begin{table}[t]
\hspace{-1cm}
\small
\begin{tabular}{ccccccccccc}
\hline
$\theta$ & $R$(mm) & $R_s$(mm) & $h_s$(mm) & Bond number & $k_R=0.2$ & $k_R=0.5$ & $k_R=1$ & $k_R=2$ & $k_R=5$ & $k_R=7$ \\
\hline
$32^\circ$ & 1 & 1.25 & 0.2  & 0.084 & single  & single  & single  & single  & single  & single  \\
$30^\circ$ & 1 & 1.25 & 0.2  & 0.083 & two     & single  & single  & single  & single  & single  \\
$30^\circ$ & 2 & 3    & 0.3  & 0.313 & two     & single  & single  & single  & single  & single  \\
$30^\circ$ & 3 & 5    & 0.5  & 0.641 & two     & two     & single  & single  & single  & single  \\
$30^\circ$ & 4 & 6    & 0.75 & 1.021 & two     & two     & two     & single  & single  & single  \\
$30^\circ$ & 5 & 8    & 1    & 1.412 & two     & two     & two     & two     & single  & single  \\
$30^\circ$ & 6 & 9    & 1    & 1.794 & two     & two     & two     & two     & two     & single  \\
$30^\circ$ & 7 & 10   & 1    & 2.157 & two     & two     & two     & two     & two     & two     \\
$30^\circ$ & 9 & 12   & 2    & 2.834 & two     & two     & two     & two     & two     & two     \\
\hline
\end{tabular}
\caption{
\label{tab:butanolL}
Numbers of vortices in the droplets of 1-butanol of different size.}
\end{table}

Domination of the evaporative cooling as compared to the heat flow
at the substrate-liquid interface leads to the reversed single-vortex 
regime for small values of $k_R$ and at small contact angles,
for weakly volatile liquids~\cite{barash2015}.
For highly volatile liquids, however, the cooling of the droplet bulk
is a faster process resulting in a larger temperature difference
between the droplet bulk and the substrate, which, in turn, 
leads to increasing the heat flow at the substrate-liquid interface
close to the contact line even for moderately low values of $k_R$.
This is in accordance with the disappearance of the subregion III
for highly volatile liquids as observed in Fig.~\ref{fig:threeliquids}.

Increasing the fluid volatility results in larger chances of domination of
the vaporization heat for relatively large contact angles.
This is in accordance with the shift of the subregion I 
at relatively small values of $k_R$ to larger contact  angles.
Understanding of the complex structure of subregion I shown in Fig.~\ref{fig:threeliquids}(c)
for relatively large values of $k_R$ requires an additional study.

Consider now the effect of droplet size on the phase diagram.
The fluid flow structure has been calculated for droplet size exceeding that
in Table~\ref{table1}.
The droplet shape becomes nonspherical
when $B_o \gtrsim 1$, where $B_o=\rho ghR/(2\sigma\sin\theta)$ 
is Bond number for the sessile droplet, where $h$ is the droplet height.
In this case, we obtained the droplet shape taking into account influence of gravity.
Nonspherical droplet free surface was obtained with the Runge-Kutta numerical method
using the approach described in~\citep{barash2009}.

We find that the phase diagrams for droplets of larger size do not contain the subregions I and II$_3$.
Table~\ref{tab:butanolL} shows numbers of vortices in the droplet of 1-butanol of different size for $\theta\approx 30^\circ$.
It demonstrates that
droplets of capillary size contain a single-vortex for $\theta\approx 30^\circ$, while larger droplets
contain two vortices in this case.
Although the existence of subregion I depends on $k_R$,
the subregion I disappears with gradually increasing the droplet radius $R$ for each $k_R$,
as shown in Table~\ref{tab:butanolL}.

The absence of subregion I for large droplets can be understood as follows.
In order for the subregion I to occur, the heat transfer through the solid-liquid interface should 
dominate compared to the vaporization heat. For droplets of larger size, the former heat flow becomes 
relatively weak for the regions far away from the substrate. Therefore, the vaporization heat dominates 
there, and the vortex splits into two vortices. 

\section{Conclusion}

We have performed numerical simulations of the fluid flow structure in evaporating
sessile droplets depending on thermal conductivities of liquid and substrate,
liquid volatility, contact angle and droplet size,
and we have analyzed the corresponding phase diagrams in the $k_R$--$\theta$ plane.

We find that with an increase of the fluid volatility,
the subregion I becomes more complex and the subregion III disappears;
the subregion II$_2$ increases at the expense of the subregion III; 
the subregion II$_3$ becomes larger at the expense of the subregion I.
Also, we have shown that the phase diagrams for droplets of larger size 
do not contain the subregions I and II$_3$.

The results may 
provide a better understanding of the Marangoni effect of drying droplets and 
may be useful for preparing the fluid flow vortex structure in a controlled manner by 
selecting substrates and liquids with appropriate properties.
Also, the results may potentially be useful to better understand
evaporative deposition patterns and evaporative self-assembly of nanostructures.

\section*{Acknowledgments}

This work was supported by the Russian Science Foundation project No. 18-71-10061.

\bibliography{refs}

\begin{thebibliography}{31}%
\makeatletter
\providecommand \@ifxundefined [1]{%
 \@ifx{#1\undefined}
}%
\providecommand \@ifnum [1]{%
 \ifnum #1\expandafter \@firstoftwo
 \else \expandafter \@secondoftwo
 \fi
}%
\providecommand \@ifx [1]{%
 \ifx #1\expandafter \@firstoftwo
 \else \expandafter \@secondoftwo
 \fi
}%
\providecommand \natexlab [1]{#1}%
\providecommand \enquote  [1]{``#1''}%
\providecommand \bibnamefont  [1]{#1}%
\providecommand \bibfnamefont [1]{#1}%
\providecommand \citenamefont [1]{#1}%
\providecommand \href@noop [0]{\@secondoftwo}%
\providecommand \href [0]{\begingroup \@sanitize@url \@href}%
\providecommand \@href[1]{\@@startlink{#1}\@@href}%
\providecommand \@@href[1]{\endgroup#1\@@endlink}%
\providecommand \@sanitize@url [0]{\catcode `\\12\catcode `\$12\catcode
  `\&12\catcode `\#12\catcode `\^12\catcode `\_12\catcode `\%12\relax}%
\providecommand \@@startlink[1]{}%
\providecommand \@@endlink[0]{}%
\providecommand \url  [0]{\begingroup\@sanitize@url \@url }%
\providecommand \@url [1]{\endgroup\@href {#1}{\urlprefix }}%
\providecommand \urlprefix  [0]{URL }%
\providecommand \Eprint [0]{\href }%
\providecommand \doibase [0]{http://dx.doi.org/}%
\providecommand \selectlanguage [0]{\@gobble}%
\providecommand \bibinfo  [0]{\@secondoftwo}%
\providecommand \bibfield  [0]{\@secondoftwo}%
\providecommand \translation [1]{[#1]}%
\providecommand \BibitemOpen [0]{}%
\providecommand \bibitemStop [0]{}%
\providecommand \bibitemNoStop [0]{.\EOS\space}%
\providecommand \EOS [0]{\spacefactor3000\relax}%
\providecommand \BibitemShut  [1]{\csname bibitem#1\endcsname}%
\let\auto@bib@innerbib\@empty
\bibitem [{\citenamefont {Brutin}\ and\ \citenamefont
  {Starov}(2018)}]{brutin2018}%
  \BibitemOpen
  \bibfield  {author} {\bibinfo {author} {\bibfnamefont {D}~\bibnamefont
  {Brutin}}\ and\ \bibinfo {author} {\bibfnamefont {V}~\bibnamefont {Starov}},\
  }\bibfield  {title} {\enquote {\bibinfo {title} {Recent advances in droplet
  wetting and evaporation},}\ }\href@noop {} {\bibfield  {journal} {\bibinfo
  {journal} {Chem. Soc. Rev.}\ }\textbf {\bibinfo {volume} {47}},\ \bibinfo
  {pages} {558--585} (\bibinfo {year} {2018})}\BibitemShut {NoStop}%
\bibitem [{\citenamefont {Erbil}(2012)}]{erbil2012}%
  \BibitemOpen
  \bibfield  {author} {\bibinfo {author} {\bibfnamefont {H~Yildirim}\
  \bibnamefont {Erbil}},\ }\bibfield  {title} {\enquote {\bibinfo {title}
  {Evaporation of pure liquid sessile and spherical suspended drops: A
  review},}\ }\href@noop {} {\bibfield  {journal} {\bibinfo  {journal} {Adv.
  Colloid Interface Sci.}\ }\textbf {\bibinfo {volume} {170}},\ \bibinfo
  {pages} {67--86} (\bibinfo {year} {2012})}\BibitemShut {NoStop}%
\bibitem [{\citenamefont {Larson}(2014)}]{larson2014}%
  \BibitemOpen
  \bibfield  {author} {\bibinfo {author} {\bibfnamefont {Ronald~G}\
  \bibnamefont {Larson}},\ }\bibfield  {title} {\enquote {\bibinfo {title}
  {Transport and deposition patterns in drying sessile droplets},}\ }\href@noop
  {} {\bibfield  {journal} {\bibinfo  {journal} {AIChE J.}\ }\textbf {\bibinfo
  {volume} {60}},\ \bibinfo {pages} {1538--1571} (\bibinfo {year}
  {2014})}\BibitemShut {NoStop}%
\bibitem [{\citenamefont {Thiele}(2014)}]{thiele2014}%
  \BibitemOpen
  \bibfield  {author} {\bibinfo {author} {\bibfnamefont {Uwe}\ \bibnamefont
  {Thiele}},\ }\bibfield  {title} {\enquote {\bibinfo {title} {Patterned
  deposition at moving contact lines},}\ }\href@noop {} {\bibfield  {journal}
  {\bibinfo  {journal} {Adv. Colloid Interface Sci.}\ }\textbf {\bibinfo
  {volume} {206}},\ \bibinfo {pages} {399--413} (\bibinfo {year}
  {2014})}\BibitemShut {NoStop}%
\bibitem [{\citenamefont {Zang}\ \emph {et~al.}(2019)\citenamefont {Zang},
  \citenamefont {Tarafdar}, \citenamefont {Tarasevich}, \citenamefont
  {Choudhury},\ and\ \citenamefont {Dutta}}]{tarasevich2019}%
  \BibitemOpen
  \bibfield  {author} {\bibinfo {author} {\bibfnamefont {Duyang}\ \bibnamefont
  {Zang}}, \bibinfo {author} {\bibfnamefont {Sujata}\ \bibnamefont {Tarafdar}},
  \bibinfo {author} {\bibfnamefont {Yuri~Yu.}\ \bibnamefont {Tarasevich}},
  \bibinfo {author} {\bibfnamefont {Moutushi~Dutta}\ \bibnamefont {Choudhury}},
  \ and\ \bibinfo {author} {\bibfnamefont {Tapati}\ \bibnamefont {Dutta}},\
  }\bibfield  {title} {\enquote {\bibinfo {title} {Evaporation of a droplet:
  From physics to applications},}\ }\href@noop {} {\bibfield  {journal}
  {\bibinfo  {journal} {Phys. Rep.}\ }\textbf {\bibinfo {volume} {804}},\
  \bibinfo {pages} {1--56} (\bibinfo {year} {2019})}\BibitemShut {NoStop}%
\bibitem [{\citenamefont {Harris}\ \emph {et~al.}(2007)\citenamefont {Harris},
  \citenamefont {Hu}, \citenamefont {Conrad},\ and\ \citenamefont
  {Lewis}}]{harris2007}%
  \BibitemOpen
  \bibfield  {author} {\bibinfo {author} {\bibfnamefont {Daniel~J.}\
  \bibnamefont {Harris}}, \bibinfo {author} {\bibfnamefont {Hua}\ \bibnamefont
  {Hu}}, \bibinfo {author} {\bibfnamefont {Jacinta~C.}\ \bibnamefont {Conrad}},
  \ and\ \bibinfo {author} {\bibfnamefont {Jennifer~A.}\ \bibnamefont
  {Lewis}},\ }\bibfield  {title} {\enquote {\bibinfo {title} {Patterning
  colloidal films via evaporative lithography},}\ }\href@noop {} {\bibfield
  {journal} {\bibinfo  {journal} {Phys. Rev. Lett.}\ }\textbf {\bibinfo
  {volume} {98}},\ \bibinfo {pages} {148301} (\bibinfo {year}
  {2007})}\BibitemShut {NoStop}%
\bibitem [{\citenamefont {Kolegov}\ and\ \citenamefont
  {Barash}(2020)}]{KolegovBarash2020}%
  \BibitemOpen
  \bibfield  {author} {\bibinfo {author} {\bibfnamefont {K.~S.}\ \bibnamefont
  {Kolegov}}\ and\ \bibinfo {author} {\bibfnamefont {L.~Yu.}\ \bibnamefont
  {Barash}},\ }\bibfield  {title} {\enquote {\bibinfo {title} {Applying
  droplets and films in evaporative lithography},}\ }\href@noop {} {\bibfield
  {journal} {\bibinfo  {journal} {Preprint 2005.07148}\ } (\bibinfo {year}
  {2020})}\BibitemShut {NoStop}%
\bibitem [{\citenamefont {Bigioni}\ \emph {et~al.}(2006)\citenamefont
  {Bigioni}, \citenamefont {Lin}, \citenamefont {Nguyen}, \citenamefont
  {Corwin}, \citenamefont {Witten},\ and\ \citenamefont
  {Jaeger}}]{bigioni2006}%
  \BibitemOpen
  \bibfield  {author} {\bibinfo {author} {\bibfnamefont {Terry~P}\ \bibnamefont
  {Bigioni}}, \bibinfo {author} {\bibfnamefont {Xiao-Min}\ \bibnamefont {Lin}},
  \bibinfo {author} {\bibfnamefont {Toan~T}\ \bibnamefont {Nguyen}}, \bibinfo
  {author} {\bibfnamefont {Eric~I}\ \bibnamefont {Corwin}}, \bibinfo {author}
  {\bibfnamefont {Thomas~A}\ \bibnamefont {Witten}}, \ and\ \bibinfo {author}
  {\bibfnamefont {Heinrich~M}\ \bibnamefont {Jaeger}},\ }\bibfield  {title}
  {\enquote {\bibinfo {title} {Kinetically driven self assembly of highly
  ordered nanoparticle monolayers},}\ }\href@noop {} {\bibfield  {journal}
  {\bibinfo  {journal} {Nat. Mater.}\ }\textbf {\bibinfo {volume} {5}},\
  \bibinfo {pages} {265} (\bibinfo {year} {2006})}\BibitemShut {NoStop}%
\bibitem [{\citenamefont {Savino}\ and\ \citenamefont
  {Fico}(2004)}]{Savino2004}%
  \BibitemOpen
  \bibfield  {author} {\bibinfo {author} {\bibfnamefont {R.}~\bibnamefont
  {Savino}}\ and\ \bibinfo {author} {\bibfnamefont {S.}~\bibnamefont {Fico}},\
  }\bibfield  {title} {\enquote {\bibinfo {title} {Transient marangoni
  convection in hanging evaporating drops},}\ }\href@noop {} {\bibfield
  {journal} {\bibinfo  {journal} {Phys. Fluids}\ }\textbf {\bibinfo {volume}
  {16}},\ \bibinfo {pages} {3738--3754} (\bibinfo {year} {2004})}\BibitemShut
  {NoStop}%
\bibitem [{\citenamefont {Kang}\ \emph {et~al.}(2004)\citenamefont {Kang},
  \citenamefont {Lee}, \citenamefont {Lee},\ and\ \citenamefont
  {Kang}}]{Kang2004}%
  \BibitemOpen
  \bibfield  {author} {\bibinfo {author} {\bibfnamefont {Kwan~Hyoung}\
  \bibnamefont {Kang}}, \bibinfo {author} {\bibfnamefont {Sang~Joon}\
  \bibnamefont {Lee}}, \bibinfo {author} {\bibfnamefont {Choung~Mook}\
  \bibnamefont {Lee}}, \ and\ \bibinfo {author} {\bibfnamefont {In~Seok}\
  \bibnamefont {Kang}},\ }\bibfield  {title} {\enquote {\bibinfo {title}
  {Quantitative visualization of flow inside an evaporating droplet using the
  ray tracing method},}\ }\href@noop {} {\bibfield  {journal} {\bibinfo
  {journal} {Meas. Sci. Technol.}\ }\textbf {\bibinfo {volume} {15}},\ \bibinfo
  {pages} {1104--1112} (\bibinfo {year} {2004})}\BibitemShut {NoStop}%
\bibitem [{\citenamefont {Hu}\ and\ \citenamefont
  {Larson}(2006)}]{HuLarsonReversal}%
  \BibitemOpen
  \bibfield  {author} {\bibinfo {author} {\bibfnamefont {Hua}\ \bibnamefont
  {Hu}}\ and\ \bibinfo {author} {\bibfnamefont {Ronald~G.}\ \bibnamefont
  {Larson}},\ }\bibfield  {title} {\enquote {\bibinfo {title} {Marangoni effect
  reverses coffee-ring depositions},}\ }\href@noop {} {\bibfield  {journal}
  {\bibinfo  {journal} {J. Phys. Chem. B}\ }\textbf {\bibinfo {volume} {110}},\
  \bibinfo {pages} {7090--7094} (\bibinfo {year} {2006})}\BibitemShut {NoStop}%
\bibitem [{\citenamefont {Ristenpart}\ \emph {et~al.}(2007)\citenamefont
  {Ristenpart}, \citenamefont {Kim}, \citenamefont {Domingues}, \citenamefont
  {Wan},\ and\ \citenamefont {Stone}}]{ristenpart2007}%
  \BibitemOpen
  \bibfield  {author} {\bibinfo {author} {\bibfnamefont {W.~D.}\ \bibnamefont
  {Ristenpart}}, \bibinfo {author} {\bibfnamefont {P.~G.}\ \bibnamefont {Kim}},
  \bibinfo {author} {\bibfnamefont {C.}~\bibnamefont {Domingues}}, \bibinfo
  {author} {\bibfnamefont {J.}~\bibnamefont {Wan}}, \ and\ \bibinfo {author}
  {\bibfnamefont {H.~A.}\ \bibnamefont {Stone}},\ }\bibfield  {title} {\enquote
  {\bibinfo {title} {Influence of substrate conductivity on circulation
  reversal in evaporating drops},}\ }\href@noop {} {\bibfield  {journal}
  {\bibinfo  {journal} {Phys. Rev. Lett.}\ }\textbf {\bibinfo {volume} {99}},\
  \bibinfo {pages} {234502} (\bibinfo {year} {2007})}\BibitemShut {NoStop}%
\bibitem [{\citenamefont {Xu}\ \emph {et~al.}(2010)\citenamefont {Xu},
  \citenamefont {Luo},\ and\ \citenamefont {Guo}}]{xu2010}%
  \BibitemOpen
  \bibfield  {author} {\bibinfo {author} {\bibfnamefont {Xuefeng}\ \bibnamefont
  {Xu}}, \bibinfo {author} {\bibfnamefont {Jianbin}\ \bibnamefont {Luo}}, \
  and\ \bibinfo {author} {\bibfnamefont {Dan}\ \bibnamefont {Guo}},\ }\bibfield
   {title} {\enquote {\bibinfo {title} {Criterion for reversal of thermal
  marangoni flow in drying drops},}\ }\href@noop {} {\bibfield  {journal}
  {\bibinfo  {journal} {Langmuir}\ }\textbf {\bibinfo {volume} {26}},\ \bibinfo
  {pages} {1918--1922} (\bibinfo {year} {2010})}\BibitemShut {NoStop}%
\bibitem [{\citenamefont {Hu}\ and\ \citenamefont
  {Larson}(2005{\natexlab{a}})}]{HuLarsonMarangoni}%
  \BibitemOpen
  \bibfield  {author} {\bibinfo {author} {\bibfnamefont {Hua}\ \bibnamefont
  {Hu}}\ and\ \bibinfo {author} {\bibfnamefont {Ronald~G.}\ \bibnamefont
  {Larson}},\ }\bibfield  {title} {\enquote {\bibinfo {title} {Analysis of the
  effects of marangoni stresses on the microflow in an evaporating sessile
  droplet},}\ }\href@noop {} {\bibfield  {journal} {\bibinfo  {journal}
  {Langmuir}\ }\textbf {\bibinfo {volume} {21}},\ \bibinfo {pages} {3972--3980}
  (\bibinfo {year} {2005}{\natexlab{a}})}\BibitemShut {NoStop}%
\bibitem [{\citenamefont {Patil}\ \emph {et~al.}(2016)\citenamefont {Patil},
  \citenamefont {Bange}, \citenamefont {Bhardwaj},\ and\ \citenamefont
  {Sharma}}]{patil2016}%
  \BibitemOpen
  \bibfield  {author} {\bibinfo {author} {\bibfnamefont {Nagesh~D.}\
  \bibnamefont {Patil}}, \bibinfo {author} {\bibfnamefont {Prathamesh~G.}\
  \bibnamefont {Bange}}, \bibinfo {author} {\bibfnamefont {Rajneesh}\
  \bibnamefont {Bhardwaj}}, \ and\ \bibinfo {author} {\bibfnamefont {Atul}\
  \bibnamefont {Sharma}},\ }\bibfield  {title} {\enquote {\bibinfo {title}
  {Effects of substrate heating and wettability on evaporation dynamics and
  deposition patterns for a sessile water droplet containing colloidal
  particles},}\ }\href@noop {} {\bibfield  {journal} {\bibinfo  {journal}
  {Langmuir}\ }\textbf {\bibinfo {volume} {32}},\ \bibinfo {pages}
  {11958--11972} (\bibinfo {year} {2016})}\BibitemShut {NoStop}%
\bibitem [{\citenamefont {Girard}\ \emph {et~al.}(2010)\citenamefont {Girard},
  \citenamefont {Antoni},\ and\ \citenamefont {Sefiane}}]{girard2010}%
  \BibitemOpen
  \bibfield  {author} {\bibinfo {author} {\bibfnamefont {Fabien}\ \bibnamefont
  {Girard}}, \bibinfo {author} {\bibfnamefont {Micka\"el}\ \bibnamefont
  {Antoni}}, \ and\ \bibinfo {author} {\bibfnamefont {Khellil}\ \bibnamefont
  {Sefiane}},\ }\bibfield  {title} {\enquote {\bibinfo {title} {Infrared
  thermography investigation of an evaporating sessile water droplet on heated
  substrates},}\ }\href@noop {} {\bibfield  {journal} {\bibinfo  {journal}
  {Langmuir}\ }\textbf {\bibinfo {volume} {26}},\ \bibinfo {pages} {4576--4580}
  (\bibinfo {year} {2010})}\BibitemShut {NoStop}%
\bibitem [{\citenamefont {Sobac}\ and\ \citenamefont
  {Brutin}(2012)}]{sobac2012}%
  \BibitemOpen
  \bibfield  {author} {\bibinfo {author} {\bibfnamefont {B.}~\bibnamefont
  {Sobac}}\ and\ \bibinfo {author} {\bibfnamefont {D.}~\bibnamefont {Brutin}},\
  }\bibfield  {title} {\enquote {\bibinfo {title} {Thermal effects of the
  substrate on water droplet evaporation},}\ }\href@noop {} {\bibfield
  {journal} {\bibinfo  {journal} {Phys. Rev. E}\ }\textbf {\bibinfo {volume}
  {86}},\ \bibinfo {pages} {021602} (\bibinfo {year} {2012})}\BibitemShut
  {NoStop}%
\bibitem [{\citenamefont {Zhang}\ \emph {et~al.}(2014)\citenamefont {Zhang},
  \citenamefont {Ma}, \citenamefont {Xu}, \citenamefont {Luo},\ and\
  \citenamefont {Guo}}]{zhang2014}%
  \BibitemOpen
  \bibfield  {author} {\bibinfo {author} {\bibfnamefont {Kai}\ \bibnamefont
  {Zhang}}, \bibinfo {author} {\bibfnamefont {Liran}\ \bibnamefont {Ma}},
  \bibinfo {author} {\bibfnamefont {Xuefeng}\ \bibnamefont {Xu}}, \bibinfo
  {author} {\bibfnamefont {Jianbin}\ \bibnamefont {Luo}}, \ and\ \bibinfo
  {author} {\bibfnamefont {Dan}\ \bibnamefont {Guo}},\ }\bibfield  {title}
  {\enquote {\bibinfo {title} {Temperature distribution along the surface of
  evaporating droplets},}\ }\href@noop {} {\bibfield  {journal} {\bibinfo
  {journal} {Phys. Rev. E}\ }\textbf {\bibinfo {volume} {89}},\ \bibinfo
  {pages} {032404} (\bibinfo {year} {2014})}\BibitemShut {NoStop}%
\bibitem [{\citenamefont {Barash}(2015)}]{barash2015}%
  \BibitemOpen
  \bibfield  {author} {\bibinfo {author} {\bibfnamefont {L.Yu.}\ \bibnamefont
  {Barash}},\ }\bibfield  {title} {\enquote {\bibinfo {title} {Dependence of
  fluid flows in an evaporating sessile droplet on the characteristics of the
  substrate},}\ }\href@noop {} {\bibfield  {journal} {\bibinfo  {journal} {Int.
  J. Heat Mass Transfer}\ }\textbf {\bibinfo {volume} {84}},\ \bibinfo {pages}
  {419--426} (\bibinfo {year} {2015})}\BibitemShut {NoStop}%
\bibitem [{\citenamefont {Sefiane}\ \emph {et~al.}(2008)\citenamefont
  {Sefiane}, \citenamefont {Moffat}, \citenamefont {Matar},\ and\ \citenamefont
  {Craster}}]{Sefiane2008}%
  \BibitemOpen
  \bibfield  {author} {\bibinfo {author} {\bibfnamefont {K.}~\bibnamefont
  {Sefiane}}, \bibinfo {author} {\bibfnamefont {J.~R.}\ \bibnamefont {Moffat}},
  \bibinfo {author} {\bibfnamefont {O.~K.}\ \bibnamefont {Matar}}, \ and\
  \bibinfo {author} {\bibfnamefont {R.~V.}\ \bibnamefont {Craster}},\
  }\bibfield  {title} {\enquote {\bibinfo {title} {Self-excited hydrothermal
  waves in evaporating sessile drops},}\ }\href@noop {} {\bibfield  {journal}
  {\bibinfo  {journal} {Appl. Phys. Lett.}\ }\textbf {\bibinfo {volume} {93}},\
  \bibinfo {pages} {074103} (\bibinfo {year} {2008})}\BibitemShut {NoStop}%
\bibitem [{\citenamefont {Carle}\ \emph {et~al.}(2012)\citenamefont {Carle},
  \citenamefont {Sobac},\ and\ \citenamefont {Brutin}}]{Carle2012}%
  \BibitemOpen
  \bibfield  {author} {\bibinfo {author} {\bibfnamefont {F.}~\bibnamefont
  {Carle}}, \bibinfo {author} {\bibfnamefont {B.}~\bibnamefont {Sobac}}, \ and\
  \bibinfo {author} {\bibfnamefont {D.}~\bibnamefont {Brutin}},\ }\bibfield
  {title} {\enquote {\bibinfo {title} {Hydrothermal waves on ethanol droplets
  evaporating under terrestrial and reduced gravity levels},}\ }\href@noop {}
  {\bibfield  {journal} {\bibinfo  {journal} {J. Fluid Mech.}\ }\textbf
  {\bibinfo {volume} {712}},\ \bibinfo {pages} {614--623} (\bibinfo {year}
  {2012})}\BibitemShut {NoStop}%
\bibitem [{\citenamefont {Zhu}\ \emph {et~al.}(2019)\citenamefont {Zhu},
  \citenamefont {Shi},\ and\ \citenamefont {Feng}}]{Shi2019}%
  \BibitemOpen
  \bibfield  {author} {\bibinfo {author} {\bibfnamefont {Ji-Long}\ \bibnamefont
  {Zhu}}, \bibinfo {author} {\bibfnamefont {Wan-Yuan}\ \bibnamefont {Shi}}, \
  and\ \bibinfo {author} {\bibfnamefont {Lin}\ \bibnamefont {Feng}},\
  }\bibfield  {title} {\enquote {\bibinfo {title} {B\'enard-marangoni
  instability in sessile droplet evaporating at constant contact angle mode on
  heated substrate},}\ }\href@noop {} {\bibfield  {journal} {\bibinfo
  {journal} {Int. J. Heat Mass Transfer}\ }\textbf {\bibinfo {volume} {134}},\
  \bibinfo {pages} {784--795} (\bibinfo {year} {2019})}\BibitemShut {NoStop}%
\bibitem [{\citenamefont {Fuchs}(1959)}]{Fuchs1959}%
  \BibitemOpen
  \bibfield  {author} {\bibinfo {author} {\bibfnamefont {Nikolai~Albertovich}\
  \bibnamefont {Fuchs}},\ }\href@noop {} {\emph {\bibinfo {title} {Evaporation
  and droplet growth in gaseous media}}}\ (\bibinfo  {publisher} {Pergamon
  Press, Oxford},\ \bibinfo {year} {1959})\BibitemShut {NoStop}%
\bibitem [{\citenamefont {Deegan}\ \emph {et~al.}(2000)\citenamefont {Deegan},
  \citenamefont {Bakajin}, \citenamefont {Dupont}, \citenamefont {Huber},
  \citenamefont {Nagel},\ and\ \citenamefont {Witten}}]{deegan2000}%
  \BibitemOpen
  \bibfield  {author} {\bibinfo {author} {\bibfnamefont {Robert~D.}\
  \bibnamefont {Deegan}}, \bibinfo {author} {\bibfnamefont {Olgica}\
  \bibnamefont {Bakajin}}, \bibinfo {author} {\bibfnamefont {Todd~F.}\
  \bibnamefont {Dupont}}, \bibinfo {author} {\bibfnamefont {Greg}\ \bibnamefont
  {Huber}}, \bibinfo {author} {\bibfnamefont {Sidney~R.}\ \bibnamefont
  {Nagel}}, \ and\ \bibinfo {author} {\bibfnamefont {Thomas~A.}\ \bibnamefont
  {Witten}},\ }\bibfield  {title} {\enquote {\bibinfo {title} {Contact line
  deposits in an evaporating drop},}\ }\href@noop {} {\bibfield  {journal}
  {\bibinfo  {journal} {Phys. Rev. E}\ }\textbf {\bibinfo {volume} {62}},\
  \bibinfo {pages} {756--765} (\bibinfo {year} {2000})}\BibitemShut {NoStop}%
\bibitem [{\citenamefont {Hu}\ and\ \citenamefont {Larson}(2002)}]{hu2002}%
  \BibitemOpen
  \bibfield  {author} {\bibinfo {author} {\bibfnamefont {Hua}\ \bibnamefont
  {Hu}}\ and\ \bibinfo {author} {\bibfnamefont {Ronald~G.}\ \bibnamefont
  {Larson}},\ }\bibfield  {title} {\enquote {\bibinfo {title} {Evaporation of a
  sessile droplet on a substrate},}\ }\href@noop {} {\bibfield  {journal}
  {\bibinfo  {journal} {J. Phys. Chem. B}\ }\textbf {\bibinfo {volume} {106}},\
  \bibinfo {pages} {1334--1344} (\bibinfo {year} {2002})}\BibitemShut {NoStop}%
\bibitem [{\citenamefont {Barash}\ \emph {et~al.}(2009)\citenamefont {Barash},
  \citenamefont {Bigioni}, \citenamefont {Vinokur},\ and\ \citenamefont
  {Shchur}}]{barash2009}%
  \BibitemOpen
  \bibfield  {author} {\bibinfo {author} {\bibfnamefont {L.~Yu.}\ \bibnamefont
  {Barash}}, \bibinfo {author} {\bibfnamefont {T.~P.}\ \bibnamefont {Bigioni}},
  \bibinfo {author} {\bibfnamefont {V.~M.}\ \bibnamefont {Vinokur}}, \ and\
  \bibinfo {author} {\bibfnamefont {L.~N.}\ \bibnamefont {Shchur}},\ }\bibfield
   {title} {\enquote {\bibinfo {title} {Evaporation and fluid dynamics of a
  sessile drop of capillary size},}\ }\href@noop {} {\bibfield  {journal}
  {\bibinfo  {journal} {Phys. Rev. E}\ }\textbf {\bibinfo {volume} {79}},\
  \bibinfo {pages} {046301} (\bibinfo {year} {2009})}\BibitemShut {NoStop}%
\bibitem [{\citenamefont {Pearson}(1958)}]{Pearson1958}%
  \BibitemOpen
  \bibfield  {author} {\bibinfo {author} {\bibfnamefont {J.~R.~A.}\
  \bibnamefont {Pearson}},\ }\bibfield  {title} {\enquote {\bibinfo {title} {On
  convection cells induced by surface tension},}\ }\href@noop {} {\bibfield
  {journal} {\bibinfo  {journal} {J. Fluid Mech.}\ }\textbf {\bibinfo {volume}
  {4}},\ \bibinfo {pages} {489--500} (\bibinfo {year} {1958})}\BibitemShut
  {NoStop}%
\bibitem [{\citenamefont {Popov}(2005)}]{Popov2005}%
  \BibitemOpen
  \bibfield  {author} {\bibinfo {author} {\bibfnamefont {Yuri~O.}\ \bibnamefont
  {Popov}},\ }\bibfield  {title} {\enquote {\bibinfo {title} {Evaporative
  deposition patterns: Spatial dimensions of the deposit},}\ }\href@noop {}
  {\bibfield  {journal} {\bibinfo  {journal} {Phys. Rev. E}\ }\textbf {\bibinfo
  {volume} {71}},\ \bibinfo {pages} {036313} (\bibinfo {year}
  {2005})}\BibitemShut {NoStop}%
\bibitem [{\citenamefont {Hu}\ and\ \citenamefont
  {Larson}(2005{\natexlab{b}})}]{HuLarson2005}%
  \BibitemOpen
  \bibfield  {author} {\bibinfo {author} {\bibfnamefont {H.}~\bibnamefont
  {Hu}}\ and\ \bibinfo {author} {\bibfnamefont {R.~G.}\ \bibnamefont
  {Larson}},\ }\bibfield  {title} {\enquote {\bibinfo {title} {Analysis of the
  microfluid flow in an evaporating sessile droplet},}\ }\href@noop {}
  {\bibfield  {journal} {\bibinfo  {journal} {Langmuir}\ }\textbf {\bibinfo
  {volume} {21}},\ \bibinfo {pages} {3963--3971} (\bibinfo {year}
  {2005}{\natexlab{b}})}\BibitemShut {NoStop}%
\bibitem [{\citenamefont {Kolegov}\ and\ \citenamefont
  {Barash}(2019)}]{KolegovBarash2019}%
  \BibitemOpen
  \bibfield  {author} {\bibinfo {author} {\bibfnamefont {K.~S.}\ \bibnamefont
  {Kolegov}}\ and\ \bibinfo {author} {\bibfnamefont {L.~Yu.}\ \bibnamefont
  {Barash}},\ }\bibfield  {title} {\enquote {\bibinfo {title} {Joint effect of
  advection, diffusion, and capillary attraction on the spatial structure of
  particle depositions from evaporating droplets},}\ }\href@noop {} {\bibfield
  {journal} {\bibinfo  {journal} {Phys. Rev. E}\ }\textbf {\bibinfo {volume}
  {100}},\ \bibinfo {pages} {033304} (\bibinfo {year} {2019})}\BibitemShut
  {NoStop}%
\bibitem [{\citenamefont {Patankar}(2018)}]{patankar}%
  \BibitemOpen
  \bibfield  {author} {\bibinfo {author} {\bibfnamefont {Suhas~V}\ \bibnamefont
  {Patankar}},\ }\href@noop {} {\emph {\bibinfo {title} {Numerical heat
  transfer and fluid flow}}}\ (\bibinfo  {publisher} {CRC Press},\ \bibinfo
  {year} {2018})\BibitemShut {NoStop}%
\end{thebibliography}%
\end{document}